\documentclass[letterpaper, 10 pt, conference]{ieeeconf}  
\usepackage{graphicx}
\usepackage{subcaption}
\usepackage{xcolor}
\usepackage{hyperref}
\usepackage{caption}
\usepackage{amsmath} 
\usepackage{marvosym} 
\usepackage{amsmath} 
\usepackage{amsfonts} 

\IEEEoverridecommandlockouts                              

\title{\LARGE \bf
DecTest: A Decentralised Testing Architecture for Improving Data Accuracy of Blockchain Oracle
}

\author{Xueying Zeng, Youquan Xian\textsuperscript{\Letter}, Chunpei Li, Zhengdong Hu, Aoxiang Zhou, Peng Liu\textsuperscript{\Letter}
\thanks{*The research was supported in part by the Guangxi Science and Technology Major Project (No. AA22068070), the National Natural Science Foundation of China (Nos. 62166004, U21A20474), Innovation Project of Guangxi Graduate Education(No. XYCSR2024098). (\textsuperscript{\Letter} corresponding author: Peng Liu and Youquan Xian).}
\thanks{$^{1}$Xueying Zeng, Youquan Xian, Chunpei Li, Zhengdong Hu, Aoxiang Zhou and Peng Liu are with School of Computer Science and Engineering, Guangxi Normal University, and Key Lab of Education Blockchain and Intelligent Technology, Ministry of Education, Guilin 541004, China.
{\tt\small liupeng@gxnu.edu.cn, xianyouquan@stu.gxnu.edu.cn}}%
}

\begin{document}

\maketitle
\thispagestyle{empty}
\pagestyle{empty}

\begin{abstract}

Blockchain technology ensures secure and trustworthy data flow between multiple participants on the chain, but interoperability of on-chain and off-chain data has always been a difficult problem that needs to be solved. To solve the problem that blockchain systems cannot access off-chain data, oracle is introduced. However, existing research mainly focuses on the consistency and integrity of data, but ignores the problem that oracle nodes may be externally attacked or provide false data for selfish motives, resulting in the unresolved problem of data accuracy. In this paper, we introduce a new Decentralized Testing architecture (DecTest) that aims to improve data accuracy. A blockchain oracle random secret testing mechanism is first proposed to enhance the monitoring and verification of nodes by introducing a dynamic anonymized question-verification committee. Based on this, a comprehensive evaluation incentive mechanism is designed to incentivize honest work performance by evaluating nodes based on their reputation scores. The simulation results show that we successfully reduced the discrete entropy value of the acquired data and the real value of the data by 61.4\%.

\end{abstract}

\section{INTRODUCTION}
Blockchain is a system that uses distributed ledger technology with features such as immutability, openness and transparency, and decentralisation~\cite{lu2019blockchain}. This technology has already been applied in several fields such as the Internet of Things~\cite{lin2022novel}, healthcare data management~\cite{madine2020blockchain}, finance~\cite{schar2021decentralized} and supply chain~\cite{powell2022garbage} and is constantly expanding into more fields. Smart contracts, as a crucial part of blockchain technology, make it possible to execute automated contracts and protocols on distributed ledgers, thus bringing greater programmability to the blockchain system~\cite{zou2019smart}. However, the inability of smart contracts to bring in data directly from the outside world is limited by their functionality and application scope.

As an intermediary platform connecting the blockchain to the outside world, oracle acts as a trusted third party, bridging the information gap between the blockchain and the outside world and making the blockchain ecosystem more resilient and practical~\cite{pasdar2023connect}. One of the biggest challenges in introducing external data into the blockchain is how to ensure the accuracy and trustworthiness of the data, which is the famous "oracle problem"~\cite{almi2023graph,al2020trustworthy}.

Some research has been devoted to solving this problem. For example, TownCrier~\cite{zhang2016town} designs a centralized solution based on the Trusted Execution Environment (TEE), which uses the Intel Software Guard Extension (SGX) to obtain data and verify the accuracy of the data through smart contracts. Once the server of the TC host is attacked, there will be a single point of failure. So ChainLink~\cite{breidenbach2021chainlink} proposed a decentralized and distributed trust model across on-chain and off-chain components to address the single point of failure, although using a lightweight on-chain aggregation mechanism to remove outliers in oracle data proposals, it is still not able to combat fraudulent or deceptive data. In addition, Taghavi et al.~\cite{taghavi2023reinforcement} proposed a method for selecting reliable and economical oracles using a reinforcement learning model that selects one or two oracles at a time for data acquisition, with the risk of targeted attacks.

Despite extensive research on the issue of data accuracy in oracle, the whole process of introducing data into the blockchain still faces the problem of oracle being attacked or providing false or deceptive data for selfish motives. With these malicious data on the chain, users will have less trust in the entire blockchain ecosystem.

In this paper, we introduce DecTest, a novel decentralized testing framework designed to improve data accuracy. DecTest features a random secret testing mechanism for blockchain oracle systems, revitalizing node monitoring and verification by using a dynamic, anonymized question-verification committee to safeguard data integrity and authenticity. Additionally, we develop an evaluation incentive mechanism that accounts for node reputation scores, offering a fair and transparent assessment of node performance. This approach encourages nodes towards greater honesty and efficiency, ultimately boosting data accuracy.

The contributions of this paper are summarized as follows:
\begin{itemize}
    \item We design a mechanism for covertly testing oracle nodes to enhance the reliability of the system by introducing a dynamic anonymization question-verification committee and a hybrid task release strategy to efficiently detect whether a node provides false data.
    \item We design a comprehensive incentive mechanism, based on the node's reputation score for in-depth evaluation, high reputation value will increase the probability of becoming a working node and the rewards will be increased, aiming to motivate oracle nodes and improve the accuracy of data.
    \item We creatively propose the Decentralised Testing (DecTest) architecture and conduct a series of experimental evaluations. The experimental results show that the proposed scheme reduces the discrete entropy value of the acquired data for the true value of the data by 61.4\% compared to the existing baseline.
\end{itemize}
The remaining sections of this paper are organized as follows. Section~\ref{related_work} discusses related work. Section~\ref{main} describes the main elements of the mechanism. Section~\ref{reputation} designs the incentive mechanism. Section~\ref{result} presents simulation results and analysis, and Section~\ref{conclusion} provides the conclusion of this paper.

\section{Related Work}
\label{related_work}
There has been a large amount of research dedicated to solving the problem of blockchain oracle access to data accuracy, divided into two main directions: reputation-based and voting-based. These solutions provide trusted mechanisms when external data is introduced to the blockchain, giving greater transparency and security to the whole system.

\subsection{Reputation-Based Oracle}
The data provided by oracle is safeguarded from tampering by designing different mechanisms based on reputation. DiOr-SGX~\cite{woo2020distributed} uses hardware protection provided by Intel Software Guard Extensions (SGX) to select reputable nodes as leaders for data transfer in a trusted and secure execution environment. Witnet~\cite{de2017witnet} selected nodes to complete RAD tasks and obtain block-producing opportunities based on their reputation values, with higher reputation indicating greater responsibilities. Pasdar et al.~\cite{pasdar2023blockchain} developed a livestock blockchain oracle (LBO) that selected nodes based on the oracle's reputation/performance metrics, addressing the issue of accessing off-chain private sensitive data. Madine et al.~\cite{madine2020blockchain} designed a reputation system to identify abnormal behavior in oracles, kicking out nodes with low scores from the oracle network. However, using hardware to provide protected data feeds incurs high costs and is relatively difficult to maintain. The use of TLS-type proofs requires the introduction of a third-party proving authority, which has the risk of reducing decentralization.

\subsection{Voting-Based Oracle}
Oracle nodes play roles such as voters and validators to cast votes. ASTRAEA~\cite{adler2018astraea} introduced multiple entity users, including validators, submitters, and voters. Submitters delivered challenges into the system, and voters adopted low-risk/low-reward strategies, while validators pursued high-risk/high-reward strategies. Deepthought~\cite{gennaro2022deepthought} combined the voting system derived from ASTRAEA with user reputation to reward the most honest users and reduce corruption risks caused by adversarial users or lazy voters. Nelatur et al.~\cite{nelaturu2020public} proposed a decentralized oracle that uses a crowdsourced voting mechanism to determine the authenticity of queries. Cai et al.~\cite{cai2022truthful} presented a scoring scheme based on peer prediction and nonlinear betting rules, aiming to extract subjective data truthfully. Gigli et al.~\cite{gigli2023decentralized} designed oracles that required consensus on specified functions and conducted voting. Vote-based oracle are only rewarded if they agree with the majority result, which can lead to herd behavior. Also, individuals or organizations holding a large number of tokens may try to manipulate the outcome of the vote to serve their interests.

The above design solutions for the oracle focus on data consistency and integrity but do not adequately consider the problem of the oracle being attacked or providing false or deceptive data for selfish motives. Currently, there is almost no mention in existing literature of testing methods to assess the accurate feedback of data from oracle nodes. Oraichain~\cite{Oraichain} proposed quality testing for AI models, allowing AI providers to earn fees only if their AI models passed these tests, primarily incentivizing AI providers to improve the accuracy of AI models. It is mentioned in~\cite{taghavi2023reinforcement} that in choosing the most valuable oracle, other oracles will test that oracle randomly, but will be vulnerable to targeted attacks or submission of spoofed data. Simultaneously, due to the public and transparent nature of blockchain, the testing process faces constraints, particularly making it challenging to evaluate the honesty of data feeding by oracle nodes.

\section{Overview}
\label{main}
 To improve the security and reliability of the oracle system, we designed a new random secret testing mechanism, DecTest, which mainly contains an incentive mechanism and testing mechanism modules.

\subsection{DecTest Architecture}
\begin{figure}
\centering
\includegraphics[width=0.5\textwidth]{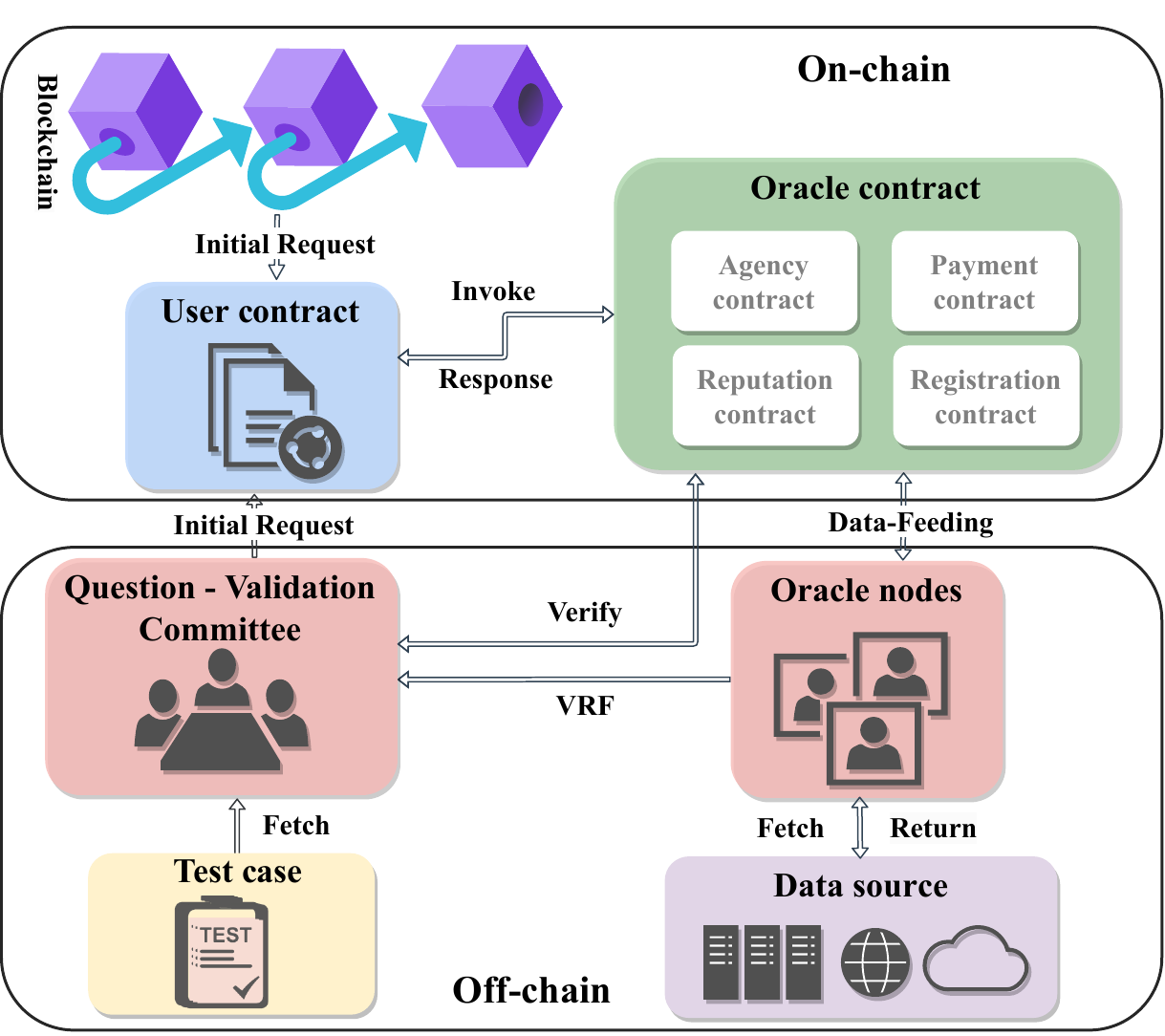}
\caption{DecTest Architecture.} \label{fig:framework}
\end{figure}

We have designed and implemented the DecTest architecture, as shown in Fig.~\ref{fig:framework}. It is mainly divided into four modules: Smart Contract, Oracle Nodes, Data Sources, and Test Cases. 

\paragraph{Smart Contract} Deploy user contracts and oracle contracts on the blockchain, with the former handling blockchain user-initiated data requests. Oracle contracts also include Registration Contracts, Payment Contracts, Reputation Contracts, and Proxy Contracts~\cite{dos}.

Registration Contract. Off-chain nodes register as oracle nodes by pledging virtual currencies, which are securely locked in the payment contract as a strong margin.

Payment Contract. The node completes the task and this deposited amount is converted into a substantial reward, providing a fair reward mechanism for the node's efforts and contributions.

Reputation Contract. At the end of each round of tasks, the reputation contract calculates and updates the reputation value of the node.

Proxy Contract. Provides an on-chain interface for user contracts and defines the API interface for oracle nodes to fetch data off-chain.

\paragraph{Oracle Nodes} Dividing the oracle into two parts, namely, the Data-Feeding Oracle and the Committee Members. The former is primarily responsible for data-feeding from the external environment to the blockchain. The latter is responsible for initiating random tests and sending task requests to the blockchain. 

\paragraph{Data Sources} The oracle gets the task and accesses the information through API to off-chain data sources.

\paragraph{Test Cases} Committee members obtain the examples needed for their test tasks from the test case repository~\cite{al2021trusting} for distribution.

\subsection{Workflow of Random Testing}
\begin{figure}
\centering
\includegraphics[width=0.5\textwidth]{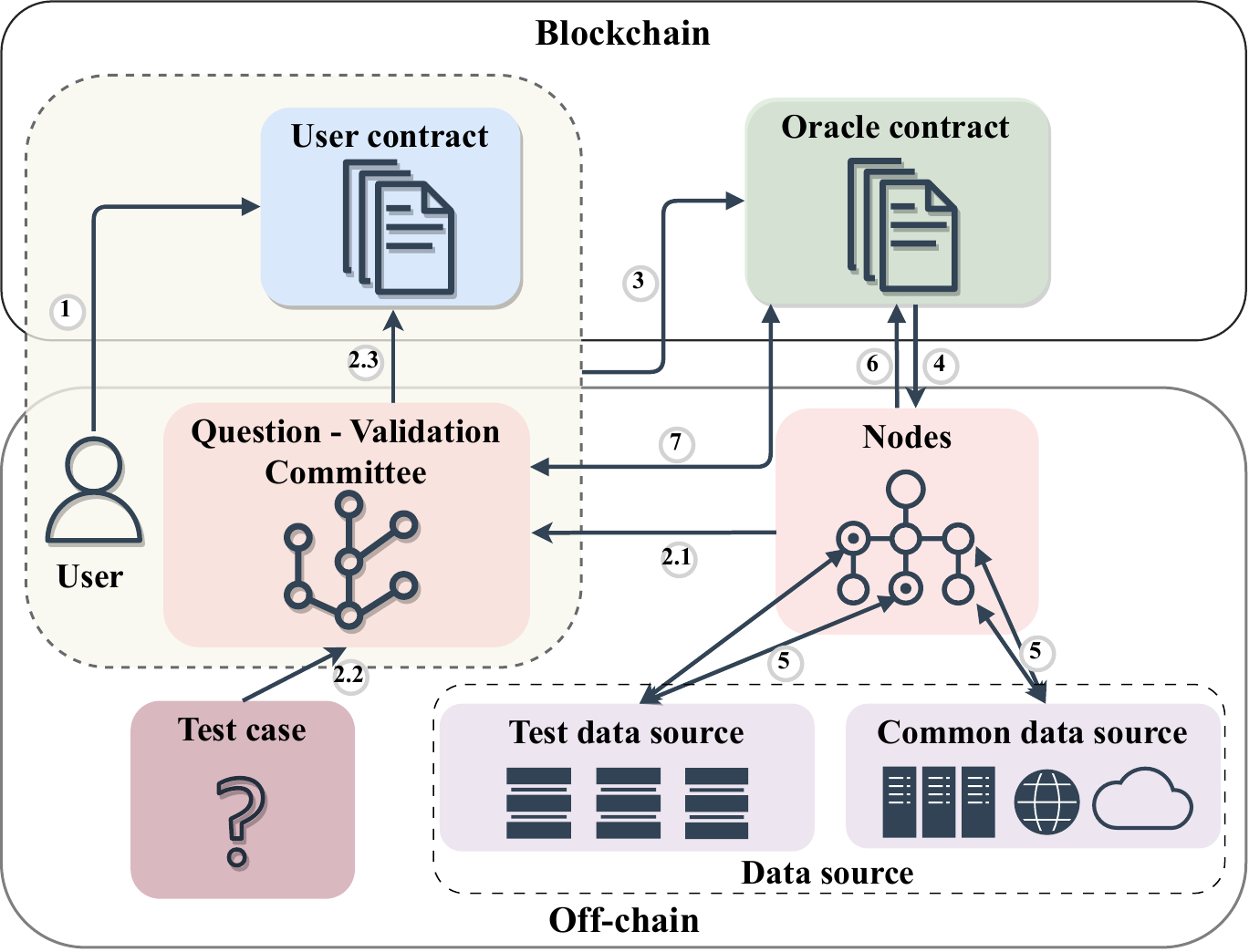}
\caption{Random secret testing process.} \label{fig:randomtesting}
\end{figure}

Fig.~\ref{fig:randomtesting} illustrates the interactions between various modules. 

\textcircled{1} When a user needs to get data, he first submits the request $Q$ to the blockchain through a smart contract.
     
\textcircled{2} The nodes are selected based on the Verifiable Random Selection Function (VRF) to form a question-verification committee, and the members generate a test data request $Q$ to be sent to the on-chain smart contract at a certain point in time. (\$III.C.1, \$III.C.2)
 
\textcircled{3} If the request is for on-chain data, the user contract directly obtains the blockchain data; if the request is for off-chain data, the predictor contract receives and responds to the request, and the blockchain will call the oracle contract after receiving the request from the user contract and passes it the relevant parameters.
 
\textcircled{4} The oracle node under the chain listens to the request $Q$ or $q$, and we design the node selection scheme based on the reputation value to select the trusted node for the under-chain data feed. (\$III.C.2)
 
\textcircled{5} The selected decentralized oracle goes to off-chain data sources such as servers, the cloud, and the internet to get data based on API.
 
\textcircled{6} The oracle node feeds the data into a smart contract on the blockchain (the node that performs the test data request also feeds the data it fetches onto the chain), and the oracle contract returns the response to the user contract and sends the response to the oracle at the same time.
 
\textcircled{7} The committee members fetch the data delivered by the oracle node from the blockchain and verify that the data fetched by the node is the same as the results of the use case, thus determining whether the oracle node is acting maliciously. (\$III.C.3) The committee feeds the results to the oracle contract to update the reputation value of the oracle node. It also calculates the rewards and punishments. (\$IV)

\subsection{Design of Random Testing}
\subsubsection{System Hypothesis}
Assuming there are $\mathcal{M}$ oracle nodes to be tested, with $\mathcal{N}$ nodes selected in each round, the testing needs to go through $\mathcal{K}$ rounds of selection and testing to achieve a coverage level with confidence $\mathcal{P}$:

\begin{equation}
\mathcal{K} = \left\lceil \frac{\ln(1-\mathcal{P})}{\ln\left(1-\frac{\mathcal{N}}{\mathcal{M}}\right)} \right\rceil
\end{equation}

We assume that blockchain can process $\mathcal {G} $ transactions per second, and $\mathcal {C} $ can process $\mathcal {X} $ test transactions per cycle. When we select $\mathcal {N} $ oracle nodes each time, we need $\mathcal N_{\mathcal {T}} $ cycles to ensure that each node has at least a probability of $\mathcal {P} $ being selected for testing:

\begin{equation}
\mathcal{N_{\mathcal{T}}} = \left \lceil \frac{\mathcal{X} \cdot \mathcal{K}}{\mathcal{G}} \cdot \frac{1}{\mathcal{C}} \right \rceil 
= \left \lceil \frac{\mathcal{X} \cdot \ln(1-\mathcal{P})}{\mathcal{G} \cdot \mathcal{C} \cdot \ln\left(1-\frac{\mathcal{N}}{\mathcal{M}}\right)} \right \rceil
\end{equation}

\subsubsection{Node Selection Strategy}
\label{choose}
To enhance the clarity of the described strategies, we can elaborate on the principles underlying the distinct methodologies employed for task oracle nodes and committee members, emphasizing the adaptive measures tailored to their unique operational contexts. This refinement involves the deployment of both a provisional registry and a blacklist list to meticulously monitor and mitigate malevolent activities within the network. Nodes exhibiting malicious intents are initially cataloged within the provisional registry; following the accrual of infractions surpassing a defined threshold, denoted by \(\theta\), such nodes are subsequently relegated to the blacklist list. 

For oracle selection in the event of an off-chain data request, we combine reputation metrics with empirical test results to ensure that nodes with demonstrable reliability are prioritized, thus enhancing the reliability of data fetching.

The formulation for the aggregation of node weights is defined as follows: Let \(sum_w\) denote the cumulative weight, calculated as \(sum_w = w_1 + \ldots + w_i \cdot \alpha + \ldots + w_n\), where \(w_i\) represents the weight of the \(i\)-th node, and \(\alpha\) signifies an adjustment factor applied to the weight of the selected node, underscoring the strategy's emphasis on enlisting nodes of higher trustworthiness to elevate the quality of data assimilated.

For the assembly of question-verification committee members, the employment of a Verifiable Random Function (VRF) is advocated to facilitate the unbiased selection of dynamically anonymous nodes, leveraging cryptographic techniques to ensure the fairness and unpredictability of the selection process~\cite{micali1999verifiable}. This method underscores the commitment to maintaining the confidentiality and integrity of the committee composition, thereby reinforcing the robustness of the verification mechanism.

\subsubsection{Structure of the question-verification committee}
\begin{figure}
\centering
\includegraphics[width=0.4\textwidth]{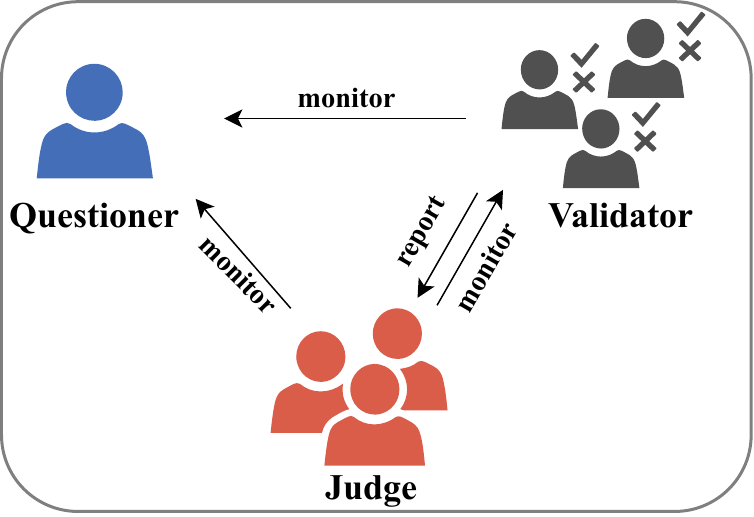}
\caption{Question-verification committee structure.} \label{fig:committee}
\end{figure}

We designed three types of nodes for the question-verification committee, as depicted in Fig. \ref{fig:committee} and described below.

Questioner. The questioner's task is to initiate a query of off-chain data. This selects a test case, publishes the task in the blockchain ecosystem, and then waits for the smart contract to execute the issued request $Q$.

Judge. Judge upon the detection of malicious endeavors, the questioner is summarily excised from the proceedings and subjected to punitive measures, concomitant with the statement of new participants. In this paradigm, the selection of a substitute questioner is predicated on a randomized algorithm, ensuring procedural integrity and fairness. Should a validator discern and report malevolent conduct by the questioner, the judge is tasked with deliberating the veracity of such claims. A determination of guilt results in the accuser assuming the role of the questioner for the ensuing cycle, whereas exoneration entails penalization of the accuser, thus maintaining a self-regulating ecosystem~\cite{zhang2022efficient}.

Validator. The validator is responsible for retrieving the blockchain and validating the data provided by oracle. The results of these validations are then transmitted to the reputation contract, which facilitates the calculation of the reputation score. This mechanism ensures continuous evaluation and validation of off-chain data, thus consolidating the reliability of the system.

To circumvent the potential for corruption and ensure the perennial integrity of the committee, a systematic rotation—denoted by the interval $cyc$—is instituted, mandating regular rejuvenation of committee membership.

\section{Incentive Mechanism}
\label{reputation}
\subsection{Reputation of Data-Feeding Oracle}
\label{oracle reputation}
We assess nodes through a reputation score design, where a higher reputation increases the likelihood of being chosen as a working node. Honest completion of data-feeding tasks results in rewards and an increase in reputation. Otherwise, nodes may face penalties. To prevent any node from becoming too influential or a target for adversarial actions due to its high reputation, we introduce a dynamic adjustment mechanism based on a threshold \(\Phi\). Once a node's reputation exceeds \(\Phi\), its reputation is moderated using an exponential decay factor, ensuring the system's resilience against targeted attacks and promoting a more equitable distribution of influence among nodes.
\begin{equation}
 \mathcal{R}_{i} = \frac{\mathcal{R}_{i}}{e^{\frac{1}{\lambda}\left ( \mathcal{R}_{i} - \Phi \right )  }}, \quad \text{for } \mathcal{R}_{i} > \Phi
\label{eq:3}
\end{equation}

We calculate the reputation score of a node based on its historical cumulative reputation $\mathcal{R}_{acci}$, response time $\mathcal{T}_{i}$, accuracy in completing random test tasks $\mathcal{AC}$, reputation weight $\mathcal{RW}$, and the number of completed test tasks $\mathcal{A}$.
\begin{equation}
\mathcal{AC}=\frac{\sum \mathcal{A}}{\sum_{i=1}^{\mathcal{K}} \mathcal{Q}_{i} }
\label{eq:7}
\end{equation}

The reputation weights not only consider the cumulative reputation of an individual relative to the mean but also introduce an adjustment factor $\Delta_{Diversity}$ to accommodate network diversity and node distribution. This ensures that nodes that contribute to network diversity are appropriately valued.
\begin{equation}
 \mathcal{RW} = \frac{\mathcal{N} \cdot \mathcal{R}_{acci}}{\sum_{i=1}^{\mathcal{N}}\mathcal{R}_{acci}} \cdot \Delta_{Diversity}
\label{eq:8}
\end{equation}

Considering the variability and consistency of node response times, a coefficient of variation $(CV)$ is introduced to reward nodes that not only respond quickly but also consistently maintain consistency.
\begin{equation}
\mathcal{RT} = \frac{CV(\{\mathcal{T}_{i}\}_{i=1}^{\mathcal{N}})}{\mathcal{N}}
\label{eq:9}
\end{equation}

The updated reputation score calculation incorporates the accuracy of task completion, reputation weight, and refined evaluation of response time. It uses a $\sigma (x)=\frac{1}{1+e^{-x} } $ function to map these components onto the (0,1) scale, ensuring that the updated scores remain within reasonable limits. $\beta ,\gamma , \delta \in \left ( 0,1 \right ) $ represents the weight ratio.

\begin{equation}
\mathcal{R}_{i}=\left ( \beta \cdot \mathcal{AC}+\gamma \cdot \sigma \left ( \mathcal{RW} \right ) +\delta \cdot \sigma \left ( \mathcal{RT} \right ) +1 \right ) \mathcal{R}_{acci}
\label{eq:10}
\end{equation}

\subsection{The Reputation of Committee Members}
\label{Committee reputation}
We introduce the concept of weighted average to more accurately reflect the variability of different members' contributions. Let \(\mathcal{R}_{i}\) represent the credibility value of the \(i\) member, and \(\omega_i\) denote the weighting factor based on his/her contribution, to distinguish the efforts and contributions of each member in a more detailed way:
\begin{equation}
\mathcal{R}_{g} = \left(\frac{\sum_{i=1}^{N} \omega_i \cdot \mathcal{R}_{i}}{\mathcal{N}}\right) \cdot \left(\frac{D}{cyc}\right)^{\rho} 
\end{equation}

\begin{equation}
\frac{D}{cyc} = \Psi(D, cyc) \quad \text{and } \quad min \leq \Psi(D, cyc) \leq max
\label{eq:12}
\end{equation}

Here, \(\rho\) is an adjustment factor to regulate the impact of the number of tasks \(D\) on rewards to better match the relationship between actual workload and rewards. This function \(\Phi\) takes into account changes in historical data, member capacity, and expected workload to dynamically adjust the number of tasks in each cycle to ensure that it stays within a reasonable range (i.e. between \(min\) and \(max\)).

\subsection{Punishment Mechanism}
\label{punish}
When we use random testing methods to detect dishonest behavior of oracle nodes that provide incorrect information, we impose appropriate penalties. This includes deducting deposit and reputation values and adding the node to the tentative list. The same penalty mechanism applies to committee members.

\begin{equation}
\mathcal{R}_{d}=\frac{\mathcal{R}_{i}}{\left( \ln(\mu + \epsilon) \right)^{d+\eta }}
\end{equation}

Here, \(d\) represents the number of times the node failed to complete the task honestly in the testing scenario, with \(\mu \in \mathbb{Z}^{*}\). \(\epsilon\) is introduced to ensure that the logarithmic function can handle very small values of \(\mu\), enhancing the formula's stability and applicability. \(\eta\) is an adjustment coefficient introduced to further modulate the penalty severity based on the level of dishonesty of the node.

When testing a node, if the number of detected instances of incorrect feedback exceeds a certain threshold \(\lambda\), the node is added to the blacklist. Concurrently, all deposited funds are deducted, and the node is removed from the oracle network.

\[
\text{If error instances} > \theta  , \text{ then execute} \left\{
     \begin{array}{l}
       \text{Add to blacklist},\\
       \text{Deduct all deposits},\\
       \text{Remove node}.
     \end{array}
   \right.
\]

\section{Experiments and Evaluation}
\label{result}
\subsection{Experiment Settings}
In this section, we conducted an experiment using 500 oracle nodes to assess the quality of data-feeding, considering the presence of a portion of malicious nodes in the system. These malicious nodes are assumed to have an 80\% probability of engaging in malicious behavior. We assume that when the number of test data sources is large enough, oracle nodes cannot determine whether they are processing a regular task or a test task through analysis. 

\begin{figure}
\centering
\includegraphics[width=0.5\textwidth]{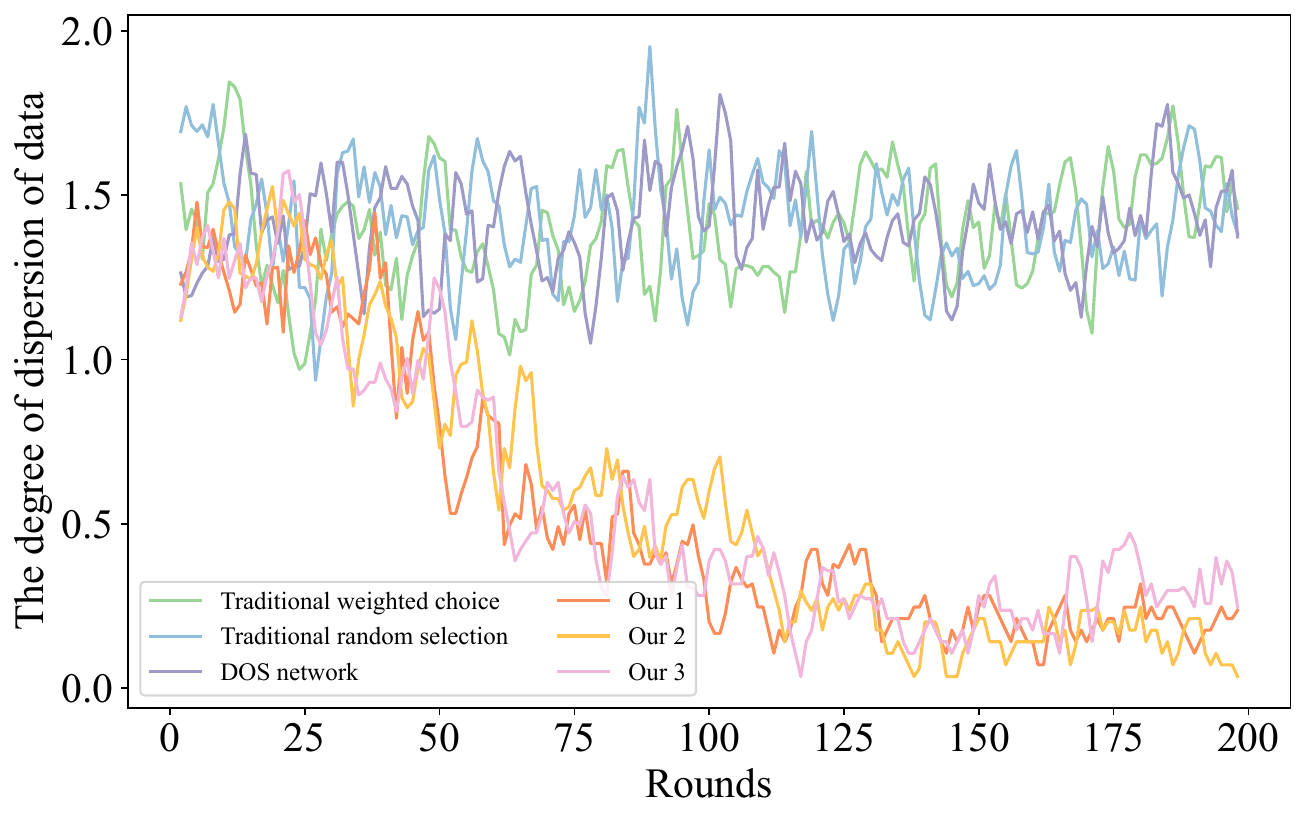}
\caption{Comparison of entropy.} \label{fig:contrast}
\end{figure}

\subsection{Evaluation Results}
To highlight the performance of adjusting node selection strategies to encourage oracle honesty based on test results, we will compare our approach with traditional weighted random node selection algorithms, pure random selection algorithms, and Dos network~\cite{dos}. The discrete entropy value of the data in the acquired data and the real value of the data was reduced by 61.4\%. Meanwhile, by adjusting the value of $\alpha$, the $\alpha$ of Our1,2,3 corresponds to [0.01,0.05], [0.1,0.3], [0.4,0.6], respectively, We found small differences in the entropy values of our methods. As shown in Fig.~\ref{fig:contrast}, we observe an overall decreasing trend in entropy values for our approach, indicating a gradual concentration of data, while other methods tend to have relatively dispersed data. This suggests that adjusting node selection weights based on test task results significantly reduces the probability of selecting malicious nodes.

\begin{figure*}[htbp]
    \centering
    \begin{subfigure}{0.32\textwidth}
        \vspace{3pt}
        \includegraphics[width=\textwidth]{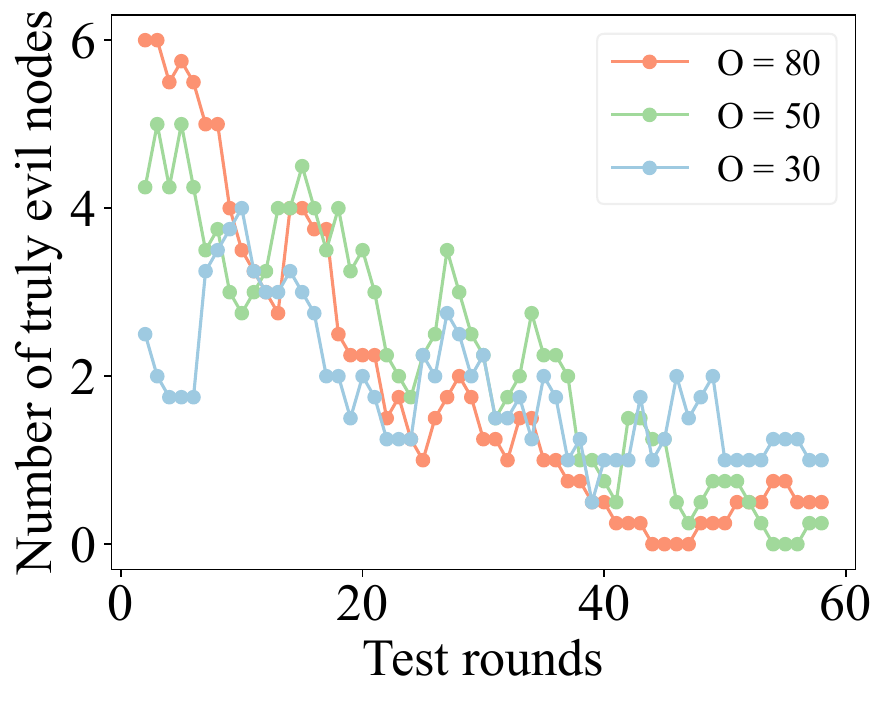}
        \caption{$s$ = 10\%}
        \label{subfig:a1}
    \end{subfigure}
    \hfill
    \begin{subfigure}{0.32\textwidth}
        \vspace{3pt}
        \includegraphics[width=\textwidth]{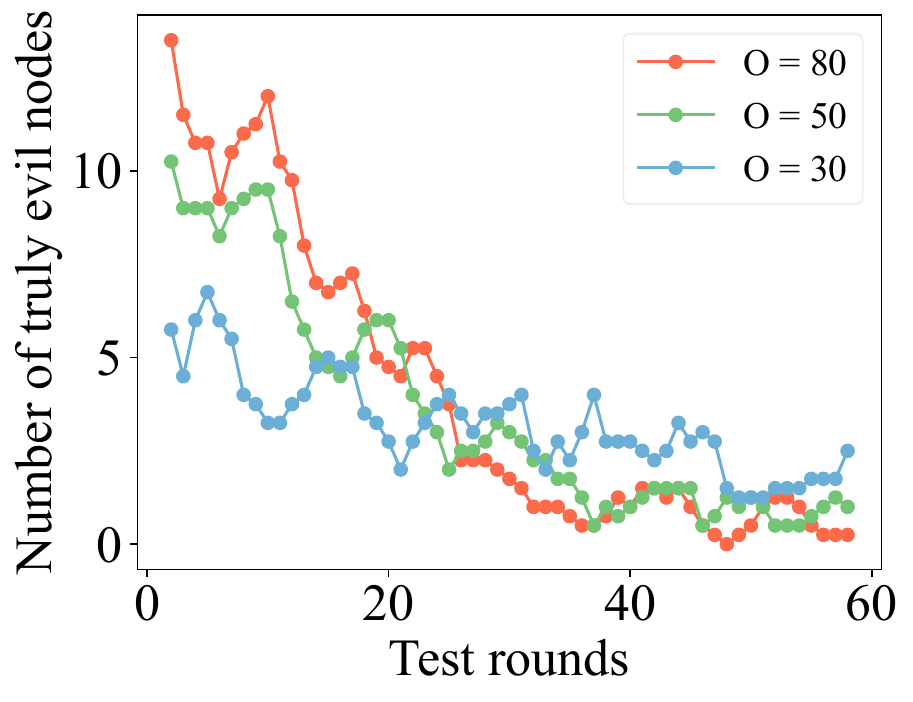}
        \caption{$s$ = 20\%}
        \label{subfig:b1}
    \end{subfigure}
    \hfill
    \begin{subfigure}{0.32\textwidth}
        \vspace{3pt}
        \includegraphics[width=\textwidth]{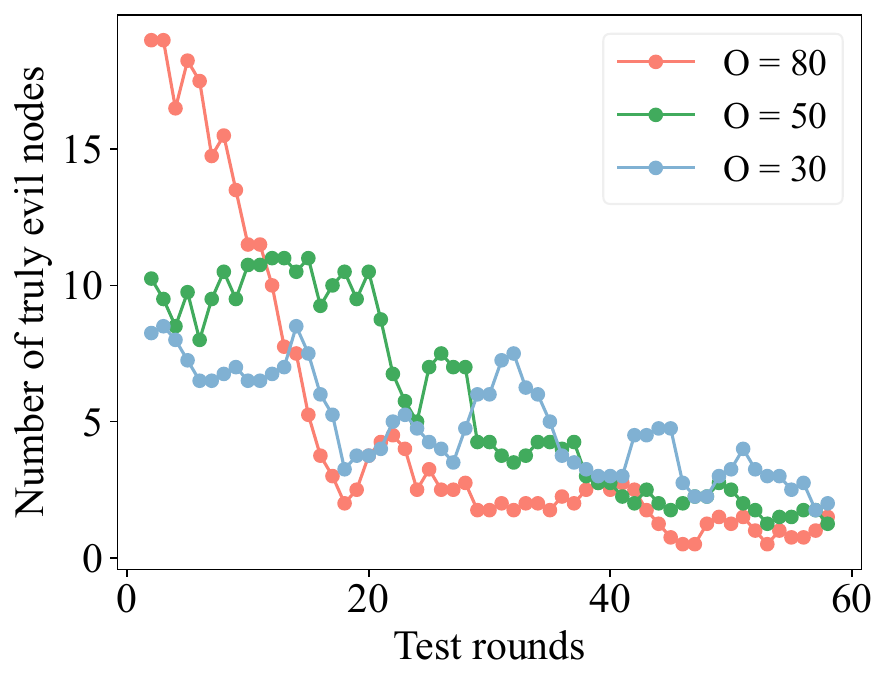}
        \caption{$s$ = 30\%}
        \label{subfig:c1}
    \end{subfigure}
 
    \caption{The real evil nodes at different percentages of malicious nodes.}
    \label{fig:capture contrast}
\end{figure*}

In order to study different percentages of malicious nodes, we conducted a series of experiments. In each round of the test task, we recorded the number of nodes with malicious behavior, as shown in Fig.~\ref{fig:capture contrast}, and the results of this experiment clearly demonstrate the process. These malicious nodes try to tamper with the data and we compare the number of truly malicious nodes accurately captured in each round by testing the mechanism under different choices of number of nodes ($n$=30, $n$=50, $n$=80). Fig.~\ref{fig:capture contrast} represents the number of nodes in the detection system that are truly evil in the case of malicious nodes $s$=10\%, 20\%, and 30\%, in the beginning, the number of malicious nodes detected is high, as the number of tests increases, the number of real malicious nodes is decreasing rapidly, it can be seen that our strategy can inhibit malicious nodes from generating malicious behavior. At the same time, with the selection of 80 decentralized oracles to complete the task, the number of truly malicious nodes is decreasing rapidly with $s$=30\% malicious nodes, although there is not a big difference between the detection of truly malicious nodes by selecting $n$=30 and $n$=50 nodes in the case of $s$=10\% malicious nodes.

\begin{figure*}[htbp]
    \centering
    \begin{subfigure}{0.32\textwidth}
        \vspace{3pt}
        \includegraphics[width=\textwidth]{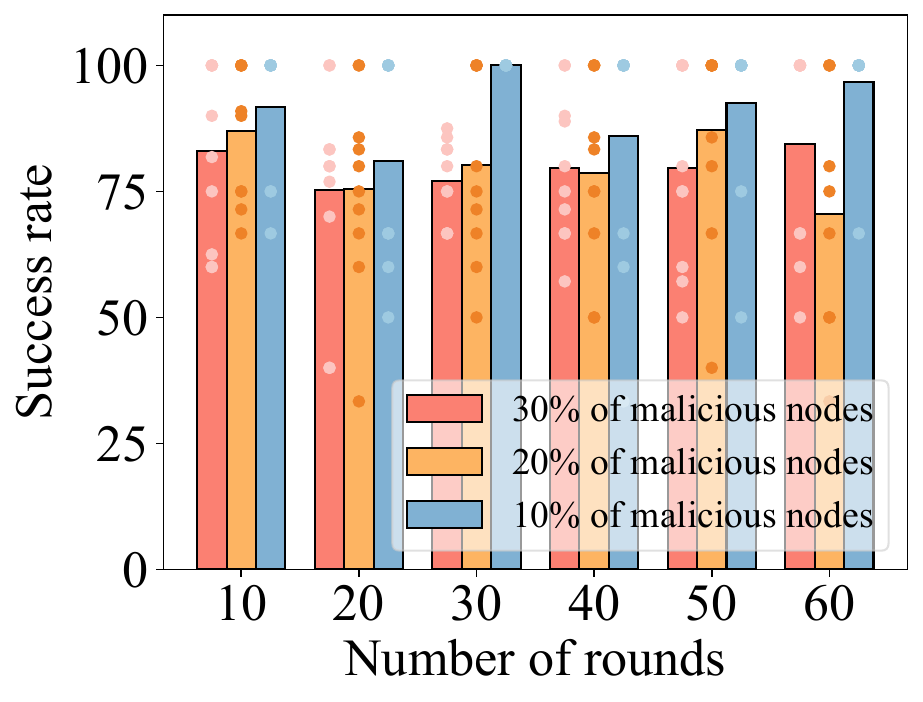}
        \caption{$n$ = 30}
        \label{fig:a2}
    \end{subfigure}
    \hfill
    \begin{subfigure}{0.32\textwidth}
        \vspace{3pt}
        \includegraphics[width=\textwidth]{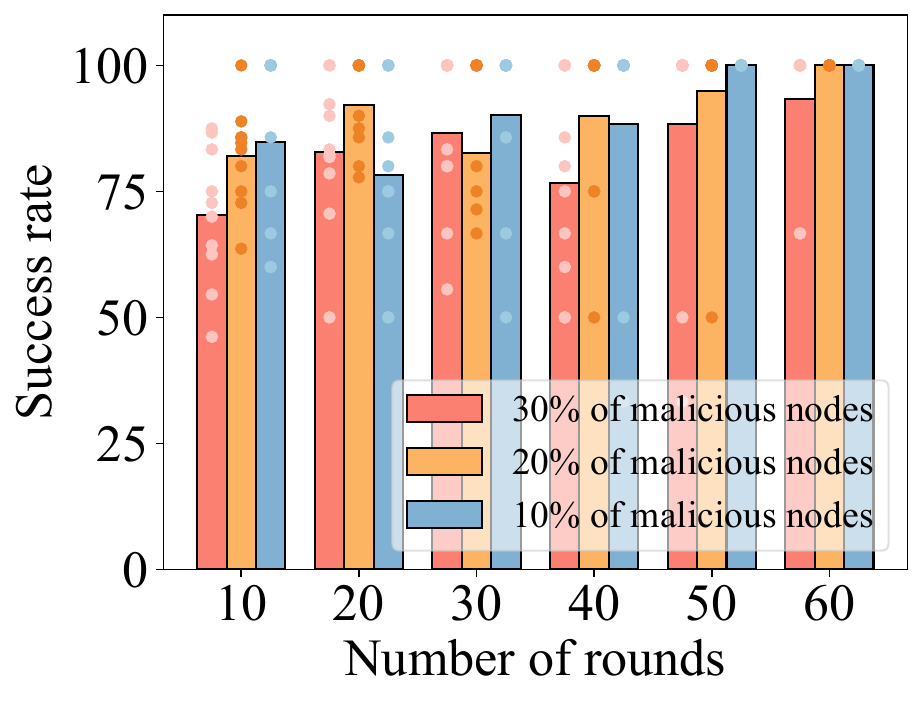}
        \caption{$n$ = 50}
        \label{fig:b2}
    \end{subfigure}
    \hfill
    \begin{subfigure}{0.32\textwidth}
        \vspace{3pt}
        \includegraphics[width=\textwidth]{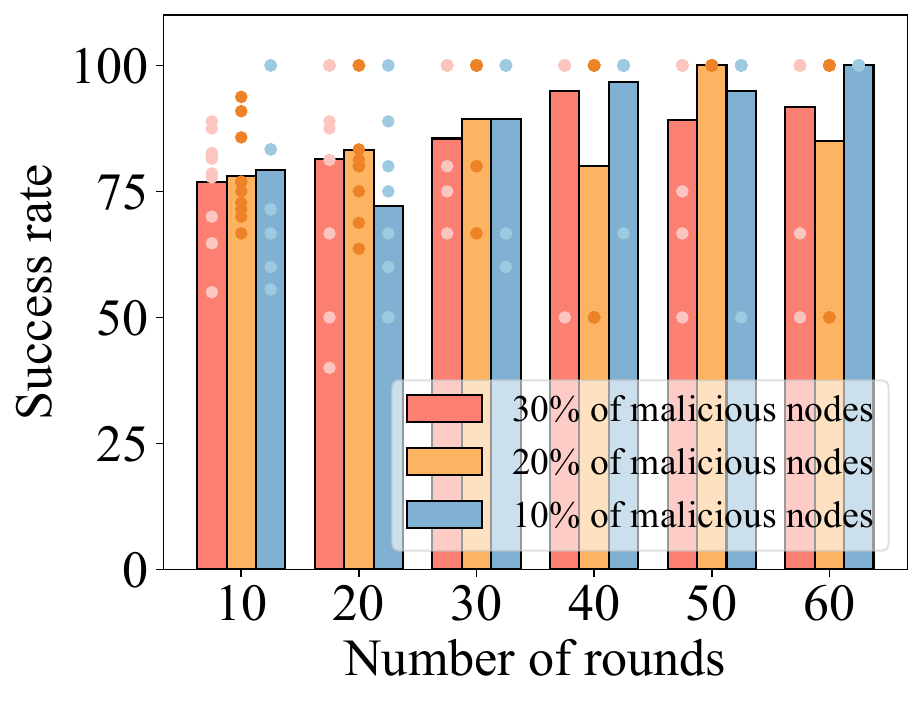}
        \caption{$n$ = 80}
        \label{fig:c2}
    \end{subfigure}

    \caption{The success rate of detecting malicious oracle nodes(bars indicate average per 10 rounds, dots indicate success rate per round).}
    \label{fig:success rate}
\end{figure*}

In addition, we also target different test rounds and different proportions of malicious nodes in order to deeply evaluate the relationship between the number of malicious nodes and the successful detection of real malicious behaviors. As shown in Fig.~\ref{fig:a2}, after selecting n=30 decentralized oracle nodes for the 60-round test task, the success rate does not show a significant increase, but it is able to achieve a success rate of 84.33\% with a malicious node percentage of $s$=30\%. In addition, Fig.~\ref{fig:b2} shows that selecting n=50 decentralized oracle nodes for the job does not show a significant improvement in the success rate of malicious node detection for each percentage in the beginning 10 rounds of testing. However, as the number of test rounds increases, the detection success rate gradually improves, and finally after 60 rounds of testing, the average detection success rate for the $s$=30\% malicious node share reaches 93.33\%. Finally, Fig.~\ref{fig:c2} shows the selection of n=80 decentralized oracle nodes to work with. Although there are some fluctuations in the success rate, the overall trend is increasing.

\begin{figure*}[htbp]
    \centering
    \begin{subfigure}{0.32\textwidth}
        \vspace{3pt}
        \includegraphics[width=\textwidth]{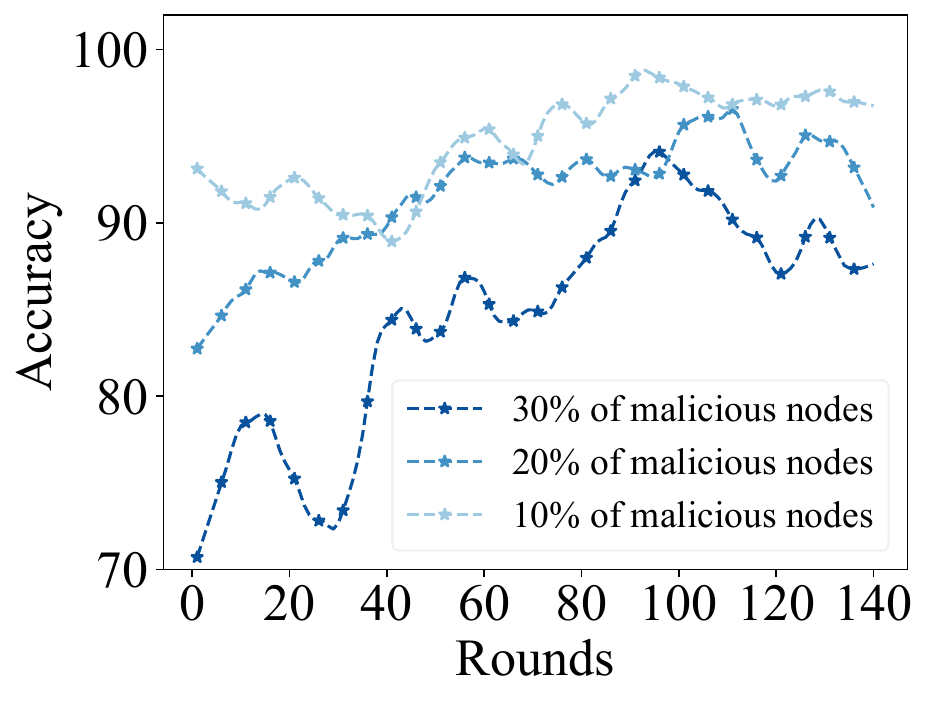}
        \label{subfig:a3}
    \end{subfigure}
    \hfill
    \begin{subfigure}{0.32\textwidth}
        \vspace{3pt}
        \includegraphics[width=\textwidth]{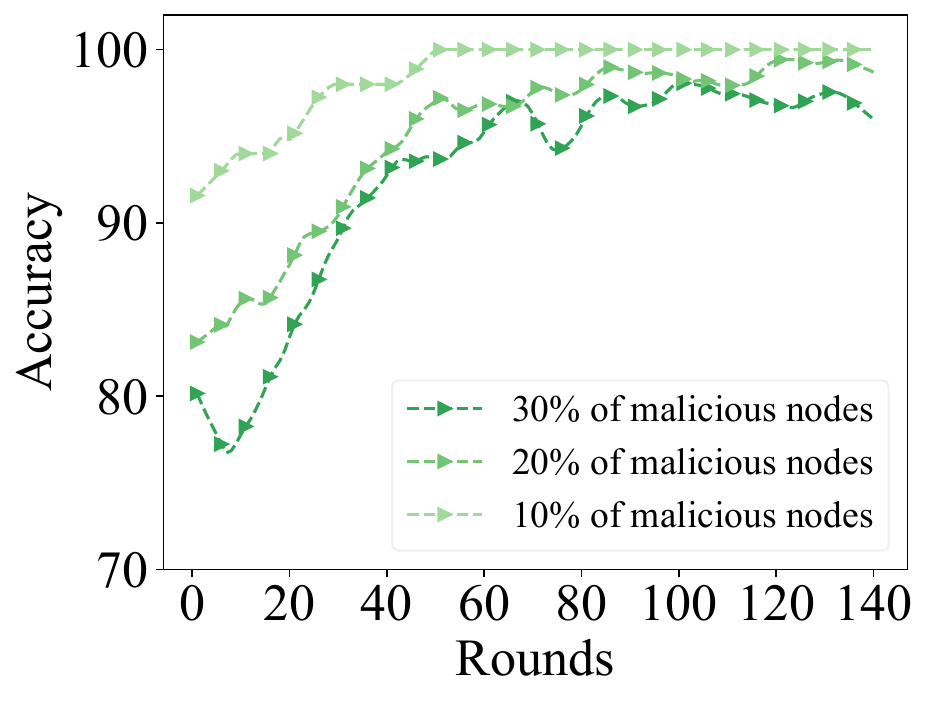}
        \label{subfig:b3}
    \end{subfigure}
    \hfill
    \begin{subfigure}{0.32\textwidth}
        \vspace{3pt}
        \includegraphics[width=\textwidth]{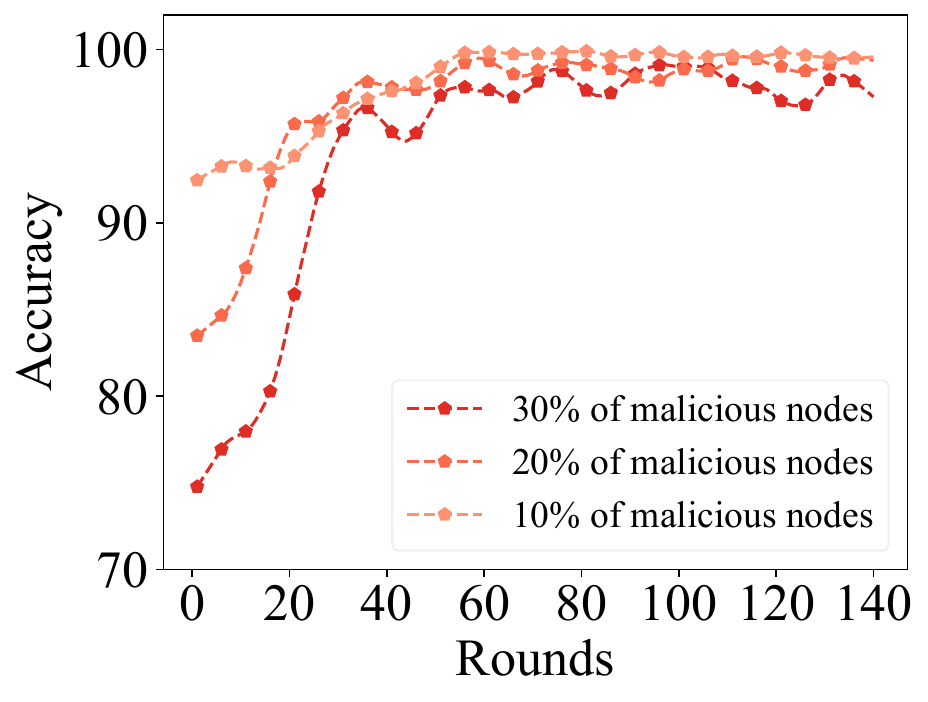}
        \label{subfig:c3}
    \end{subfigure}

    \caption{Accuracy of data-feeding(Select the number of oracle-feeding $n$ = 30, 50, 80).}
    \label{fig:accuracy rate}
\end{figure*}

Meanwhile, we also keep statistics on the accuracy of the data, and Fig.~\ref{fig:accuracy rate} demonstrates the selection of different numbers of decentralized oracle nodes to complete the data feed task. The accuracy rate of the data performs quite consistently in scenarios where the percentage of malicious nodes reaches 10\%. Regardless of the number of nodes we chose to perform the task, the accuracy rate almost stays above 90\%. In the other two scenarios with a malicious node share, the accuracy rate is not as good at first. However, it is worth noting that the accuracy rate in both cases shows a rapid increase as time passes and the number of nodes increases. This suggests that our designed system is able to gradually improve the accuracy of the data despite the possible influence of malicious nodes in the initial phase.

\section{Conclusion}
\label{conclusion}
To address the problem of oracle being subject to external attacks or providing false data for selfish motives, we propose a covert testing mechanism for oracle. The mechanism consists of randomly publishing test tasks to monitor the operation status of oracle. In addition, we design an incentive mechanism to ensure that only the oracle that works honestly will get the maximum benefit. Through simulation experiments, we have successfully verified the effectiveness of the strategy in detecting and preventing mischief. In the future, we plan to investigate more efficient detection methods, and at the same time, we will explore the application of real-world data to further improve the credibility of the prognosticator and the quality of data feeds.

\bibliographystyle{IEEEtran}
\bibliography{mybibliography}

\end{document}